# Analysis of the mechanical performance of the 4.2 m long MQXFA magnets for the Hi-Lumi LHC Upgrade

L. Garcia Fajardo, G. Ambrosio, A. Ben Yahia, D.W. Cheng, P. Ferracin, J. Ferradas Troitino, S. Izquierdo Bermudez, J. Muratore, S. Prestemon, K.L. Ray, M. Solis, G. Vallone

*Abstract*— Under the U.S. High Luminosity LHC Accelerator Upgrade Project (HL-LHC AUP), the 150 mm bore, high-field Nb₃Sn low-β MQXFA quadrupole magnets are being fabricated, assembled and tested, in the context of the CERN Hi-Luminosity LHC (HL-LHC) upgrade. These magnets have 4.2 m magnetic length and 4.56 m long iron yoke. To date, eight MQXFA magnets have been tested. One of the magnets additionally underwent a successful endurance test with 40 triggered quenches, and two magnets did not perform as expected. This work summarizes for the first time the available strain gauge data from eight identical Nb₃Sn MQXFA tested magnets, focusing on the endurance test, and on a possible cause of underperformance of the two magnets that did not pass the vertical test. We applied methods to prevent this from happening in future MQXFA magnets, which shown to be effective for last two tested magnets.

*Index Terms*—Large Hadron Collider, strain measurement, stress measurement, superconducting magnets

## I. INTRODUCTION

IN the framework of the High Luminosity upgrade of the Large Hadron Collider (HL-LHC) [1, 2], 150 mm bore, high field Nb₃Sn low-β quadrupole magnets are planned to be installed in the interaction regions of the LHC by 2026 [1]. These are called MQXF, and will be fabricated in two different magnetic lengths: the 4.2 m long MQXFA magnets, which are being fabricated in the U.S. under the U.S. HL-LHC Accelerator Upgrade Project (AUP) [3], and the 7.15 m long MQXFB magnets [4], fabricated by CERN. Both magnets have the same cross section (Fig. 1).

The MQXFA magnets are assembled at the Lawrence Berkeley National Laboratory (LBNL). Each magnet consists of four 4.53 m long coils surrounded by laminations of aluminum collars and ARMCO iron pads that are bolted together. This is called the coil pack assembly. The coils are located in the four quadrants of the coil pack. Looking from the lead end of the magnet, they are designated counterclockwise as Q1 through Q4, where Q1 corresponds to the coil located at the upper-right quadrant.

The 4.56 m long shell-yoke assembly consists of four laminated iron subassemblies placed inside the segmented aluminum shell, which consist of eight 7075 aluminum cylinders (Fig. 2). About 10 MPa initial azimuthal prestress is applied to the shells by interference keys and shims between the yoke subassemblies. The key insertion is done using bladders (welded stainless-steel sheets) between aluminum cylinders split longitudinally in half, placed in the cooling holes. The bladders are pressurized with water increasing the gaps between the iron yoke parts, allowing the insertion of the key-shim pack.

The coil pack is then inserted inside the shell-yoke assembly and the target azimuthal prestress at room temperature is applied to the coils. This is called the magnet preload process, which is done with similar bladders placed in the bladder slots and the interference key-shim pack in the load key slots (Fig. 1) [5, 6]. The key-shim packs previously placed between the iron yoke subassemblies are then removed in order to allow the shell to complete the azimuthal preload to the coils during cooldown. There are also four 31.8 mm diameter stainless steel 316 L rods extended through the magnet between the iron pads (Fig. 1) which are attached to end plates on both ends. During the magnet preload process, a piston pushes the end plate located at the return end of the magnet, applying axial prestress to the coils by stretching the four rods. The nuts on the ends are then tightened and the piston pressure is released, leaving the rods under tension.

Preload targets at room temperature are established for the coil azimuthal stress as -80 ±8 MPa, for the shell azimuthal

Manuscript receipt and acceptance dates will be inserted here. This work was supported in part by the U.S. Department of Energy, Office of Science, Office of High Energy Physics, through the US HL-LHC Accelerator Upgrade Project, and in part by the High Luminosity LHC project at CERN *(Corresponding author: Laura Garcia Fajardo)*

L. Garcia Fajardo (lgarciafajardo@lbl.gov), D. W. Cheng (dwcheng@lbl.gov), S. Prestemon (soprestemon@lbl.gov), K.L. Ray (klray@lbl.gov), M. Solis (msolis@lbl.gov), and G. Vallone (gvallone@lbl.gov) are with the Engineering Division, Lawrence Berkeley National Laboratory, Berkeley, CA 94720 USA.

P. Ferracin (pferracin@lbl.gov) is with the Accelerator Technology and Applied Physics Division, Lawrence Berkeley National Laboratory, Berkeley, CA 94720 USA.

G. Ambrosio (giorgioa@fnal.gov) is with the Fermi National Accelerator Laboratory, Batavia, IL 60510 USA.

A. Ben Yahia (abenyahia@bnl.gov) and J. Muratore (muratore@bnl.gov) are with Brookhaven National Laboratory, Upton, NY 11973 USA.

J. Ferradas Troitino (jose.ferradas.troitino@cern.ch) and S. Izquierdo Bermudez (susana.izquierdo.bermudez@cern.ch) are with the TE-MSC, CERN, 1211 Geneva, Switzerland.

Color versions of one or more of the figures in this paper are available online at http://ieeexplore.ieee.org.

Digital Object Identifier will be inserted here upon acceptance.

stress as 58 ±6 MPa, and for the rod axial strain as 950 ±95 $\mu\varepsilon$. The coils must not exceed 110 MPa azimuthal compression stress during the assembly process [6]. The MQXFA magnets are then sent to Brookhaven National Laboratory (BNL) to undergo the vertical test in liquid helium (LHe).

Eight MQXFA magnets have been assembled and tested so far: MQXFA03 trough MQXFA11, except MQXFA09 (sometimes referred in this document as A03 through A11 for abbreviation). Magnet A05 additionally underwent an endurance test in April 2022 with two thermal cycles and 40 triggered quenches. Magnets A07 and A08 did not pass the vertical test, and the limiting coil was Q3 in both cases. A08 was disassembled and rebuilt into A08b, where only the limiting coil was replaced by a new coil. A08b has not been tested yet as of this writing. A09 experienced several Non-Conformities during assembly: coil Q4 reached 121 MPa compression during the magnet assembly process, and afterwards, folded ground plane insulation was observed between coils Q2 and Q3, where subsequent analyses indicated coil conductors may have been damaged, so the magnet was disassembled. A09 structure parts were rebuilt into A11 with four new coils.

This work summarizes the available strain gauge (SG) data from the tested magnets, focusing on the A05 endurance test, and on a possible cause of underperformance of A07 and A08. We applied methods to prevent this from happening in future MQXFA magnets, which shown to be effective for A10 and A11. This is the first time we can compare the mechanical behavior of eight identical Nb$_3$Sn magnets tested in LHe.

## II. STRAIN GAUGE SETUP

The strain applied to the coils, shells and rods is monitored at every step during the assembly and during the vertical test by means of HBM SGs.

The coil SGs are 350 Ω circuits connected in half bridge configuration with two resistors: one resistor (the active gauge) is bonded to the titanium pole of the coil, and the other resistor (the passive gauge) is bonded to a small titanium plate that remains floating near the active gauge for temperature compensation. The coil SGs are placed at 3989 mm from the lead end, which is representative of the peak stress on the conductor [7, 8], in the azimuthal and axial directions.

The shell gauges have the same characteristics as the coils', and are located on shell 2, shell 4 and shell 7 (Fig. 2), on four azimuthal locations designated counterclockwise as "top", "left", "bottom" and "right" when looking at the lead end of the magnet (Fig 1). In this case, the passive gauge is bonded to a small aluminum plate.

The rod's SGs are 350 Ω circuits connected in full bridge configuration with four resistors measuring only axial strain, with two opposing gauges oriented axially, and the other two oriented azimuthally on the rod.

The stress is calculated from the strain measurements using the following equation for the coils and the shells:
$\sigma_{T,Z} = E \cdot 1e^{-3}(\varepsilon_{T,Z} + \nu \cdot \varepsilon_{Z,T})/(1-\nu^2)$, and for the rods: $\sigma_Z = E \cdot 1e^{-3} \cdot \varepsilon_Z$, where $\sigma_T$ and $\sigma_Z$ are the azimuthal and axial stress components (in MPa), respectively, $\varepsilon_T$ and $\varepsilon_Z$ are the

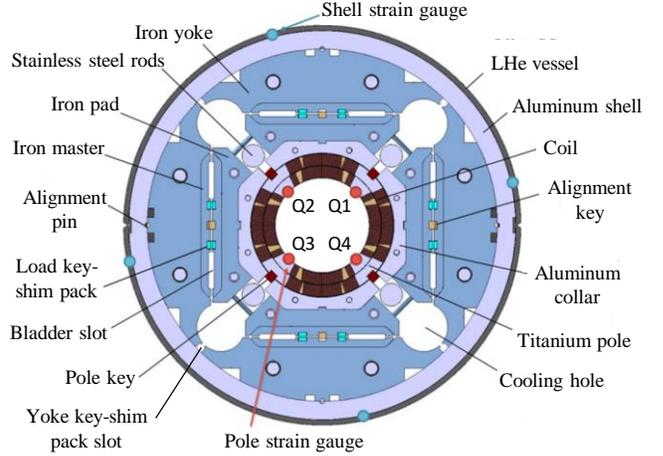
Fig. 1. Cross section of the MQXFA magnets and strain gauge location (looking at the lead-end of the magnet).

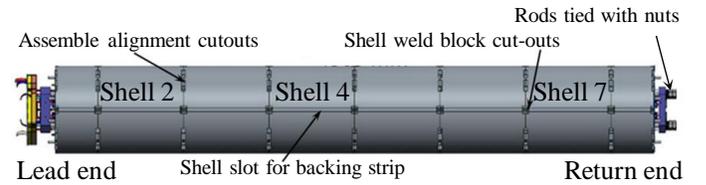
Fig. 2. Longitudinal view of the MQXFA magnets.

azimuthal and axial strain components (in $\mu\varepsilon$), respectively. $\nu$ is the Poisson's ratio (0.3), and $E$ is the Young's Modulus (in GPa). The $E$ values of titanium, aluminum and stainless steel are considered 130 GPa, 70 GPa and 195 GPa at room temperature, respectively, and 130 GPa, 79 GPa and 210 GPa at 4.2 K.

## III. MECHANICAL PERFORMANCE OF A05

Magnet A05 successfully passed the first vertical test, which ended in May 2021. It consisted of three thermal cycles from room temperature to 1.9 K. The magnet reached the acceptance current of 16.53 kA after eight quenches. The SG cooldown and warmup data from the second and third thermal cycles were not recorded, therefore, the final strain/stress state of the magnet components was undetermined. However, the first warmup data showed that the coils, shells and rods returned to their previous room temperature state.

An endurance test was done one year later: it ended in May 2022. It consisted of two thermal cycles with 40 triggered quenches and seven power ramps without quenching. The goal of this test was to demonstrate that MQXF magnets meet requirements after 5 thermal cycles and 50 quenches (HL-LHC operational scenario).

Fig. 3 shows the history of coil azimuthal stress, shell 2 azimuthal stress and rod axial strain during the endurance test. Shell 2 was chosen to represent the azimuthal behavior of the whole shell in this plot. The first point in the plots correspond to the initial SG readings before the first vertical test, and the second point represents the SG readings at the end of the first warmup from the first test. These are reference points of the previous state of the magnet for the endurance test.

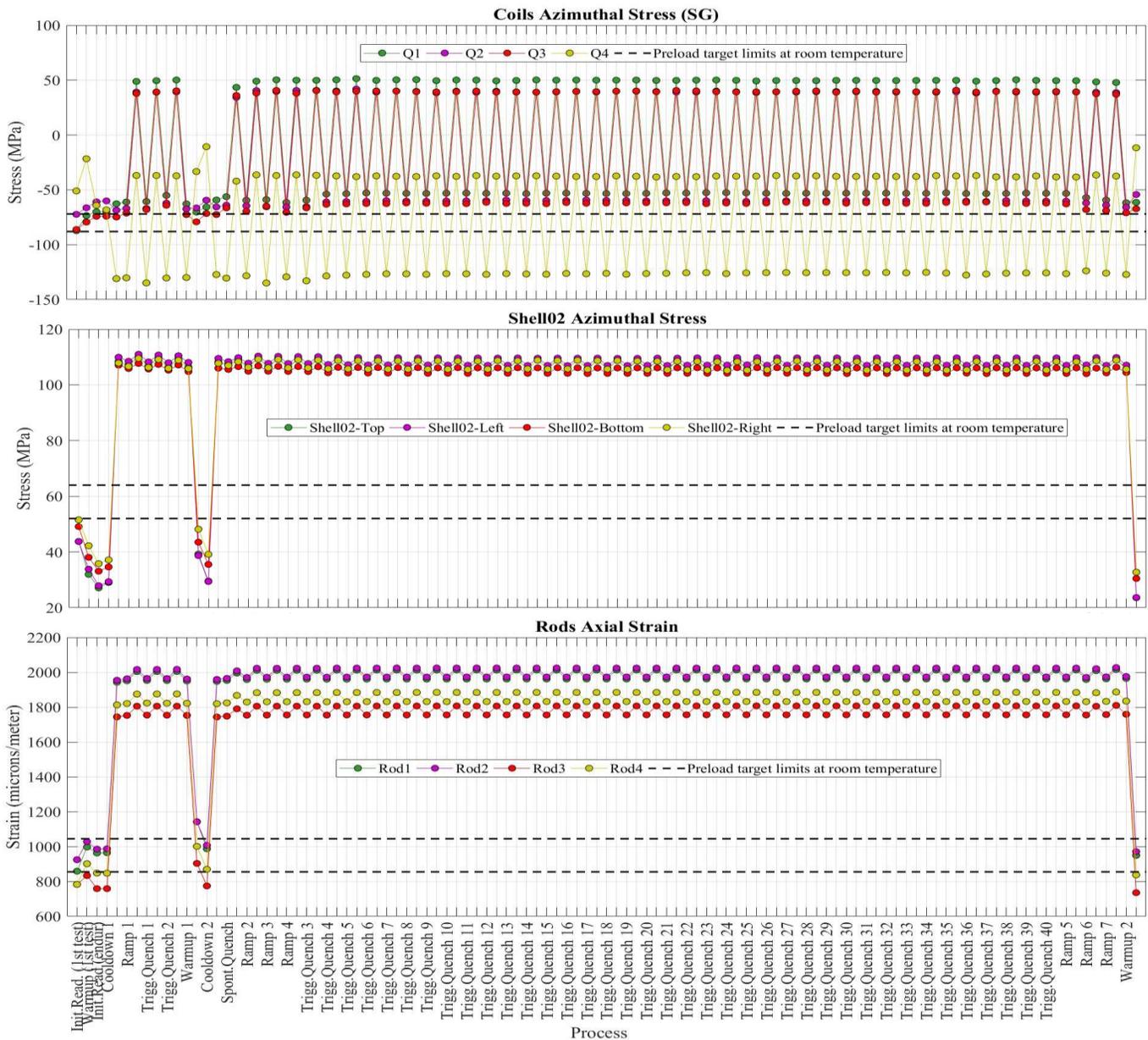

Fig. 3. History of the A05 coil and shell 2 azimuthal stress, and rod axial strain during the endurance test.

After the preload process at room temperature, the prestress of Q4 appeared to be below the target compression values (see first point in the plot). It was decided not to increase the magnet prestress because Q1 and Q3 were at the upper prestress limit, and if pushing the preload further, Q1 and Q3 would have been at risk of experiencing an excessively high compression. During cooldown from room temperature to 4.2 K, the Q4 SG signal showed very high compression, whereas minimum change, more consistent with finite element analysis [9] was observed in the SG signals from the other three coils, and has also been observed in most of the magnets tested so far. In spite of the offset observed in the Q4 data during cooldown, its delta during powering is extremely reproducible, indicating a very stable mechanical behavior. We conclude that the absolute strain values of Q4 after the preload process and during the cooldown are inaccurate. We expect that in future magnets, starting from A12, FBG sensor data will be available during the test and will help clarify this type of issue.

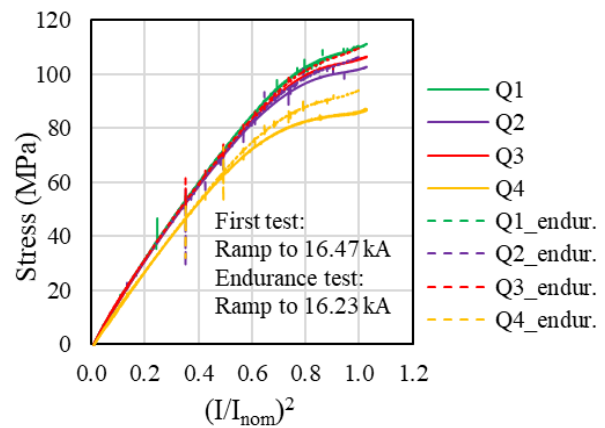

Fig. 4. Coil azimuthal stress-increase (delta stress) of A05 during excitation, during the first test (solid lines) and the endurance test (dashed lines).

The plots in Fig. 3 show a very consistent behavior of the components, especially the rods and shell 2, which show no signs of plastic deformation. This suggests that the desired level of prestress to the coils will be maintained for the magnet's lifetime.

Fig. 4 shows the coil azimuthal stress-increase (delta stress caused by decreasing compression at the coil-pole interface) when powering the magnet during the first test and the endurance test. The powering scheme consists of a ramp up to the nominal current, hold for 30 minutes, and a ramp down with the same ramp rate. The ramp-up and ramp-down legs are then averaged in order to remove the inductive components in the SG signals caused by dI/dt when the ramp rate changes.

In both tests, the coils experience a similar behavior. It is noticeable how Q4, which presented the lowest preload at room temperature, also experiences the lowest preload during excitation. The deviation of the curves from a straight line (i.e. the flattening of the curve) is a sign of partial debonding of the epoxy-impregnated coils from the titanium poles [8].

## IV. MECHANICAL PERFORMANCE OF A07 THROUGH A11

Magnets A07 and A08 were tested in August 2021 and from October 2021 to February 2022, respectively. They did not pass the vertical test, with Q3 the limiting coil in both cases. Analyses showed the quenches occurred close to the wedge spacer transition at the lead end of the magnet; a detailed analysis describing the possible causes of failure can be found in [10]. Since the SGs are located at the opposite end from where the quenches took place, it is very difficult to see any indication of the quench cause in the SG data. The only anomalous behavior observed in Q3 was a larger axial strain increase during excitation compared to the other coils. However, this is not necessarily reflected in the behavior of the rods. In addition, magnets A05 and A06 also showed a large axial strain increase in Q3, but these magnets still passed the acceptance test criteria. Therefore, from the SG data alone it is not possible to arrive to a conclusion regarding the cause of the magnet limitations.

After disassembling these two magnets, it was found out that the gap between the collars and the pole key in Q3 was closed after testing (it should remain open at room temperature). FEA models showed this condition can cause the azimuthal stress from the shell to not transfer to the titanium pole and to the coil Q3 during cooldown. As a consequence, the coil ended up less constrained axially by the support structure, and developed high axial strain in the end region [10].

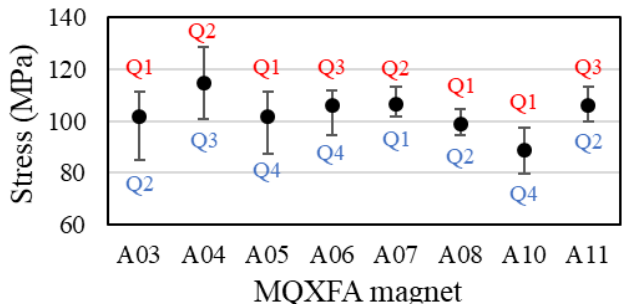

Fig. 5. Coil average azimuthal stress-increase (delta stress) during the excitation of all the magnets tested so far. The top (red) and bottom (blue) labels represent the coil with the highest and the lowest stress-increase.

This issue was assessed by revising the methodology for ensuring the squareness of the coil pack assembly in order to maintain uniform pole key gaps in the coil pack [11]. The new methodology was then applied to magnets A10 and A11 and both magnets passed the vertical test. The test of A10 ended in August 2021 after two thermal cycles with 19 quenches. The test of A11 ended in October 2022 after three thermal cycles with 15 quenches.

Fig. 5 shows the coil average azimuthal stress-increase (delta stress) in magnets A03 through A11 during excitation. The coil average stress is calculated selecting the maximum stress-increase of each coil during the ramp (ramp-up and ramp-down legs are averaged as previously described). Q3 and Q4 were not included in the calculation for A03 and A11, respectively, since the SG signal of these channels was lost. Since A07 did not reach the nominal current, the current of A07 in this plot is 15.45 kA. For the other magnets the current ranges from 16.45 to 16.65 kA; this is because before September 2020 the nominal current was 16.47 kA, and it was subsequently reduced to 16.23 kA.

The average compression stress-increase oscillates about 100 ±10 MPa. The spread about the average is also shown by means of "error" bars. The coil with the lowest (in red) and the highest (in blue) stress-increase is also shown for each magnet.

## CONCLUSIONS

The paper summarized the mechanical behavior of tested MQXFA magnets according to the SG measurements.

A05 showed an outstanding performance after a total of five thermal cycles, 52 quenches and 79 power cycles.

The absolute stress value of the coils at cold is not always reliable as observed in the coil Q4 of A05. However, if we assume that the unloading during excitation is a good representation of the coil prestress at 1.9 K, as suggested by FE models, we can conclude that the coils have a compression after cooldown of about -100 ±10 MPa based on these deltas.

The SG data cannot explain the underperformance of A07 and A08; this might be a consequence of the long distance between the SG location (return end) and the quench location (lead end) in both magnets. New assembly and measurement methods were applied to A10 and A11 to control the squareness of the coil pack, and these magnets passed the vertical test.

## ACKNOWLEDGMENT

The authors from LBNL would like to thank the technicians Ahmet Pekedis, Joshua Herrera, Juan Rodriguez, Jonathan Hekmat, and Matthew Reynolds for their outstanding job and very valuable help in the instrumentation and the assembly of the magnet parts.